\documentclass[copyright,creativecommons]{eptcs}
\providecommand{\event}{GandALF 2013} 

\usepackage{latexsym,amssymb}
\usepackage{tikz}

\def\R{\mathbb R}

\def\si{\sigma}

\def\cC{\mathcal C}             \def\cD{\mathcal D}
\def\cE{\mathcal E}             \def\cF{\mathcal F}

             \def\cP{\mathcal P}

             \def\cV{\mathcal V}
\def\cW{\mathcal W}

\def\bA{\mathbf A}              
              
              \def\bF{\mathbf F}
\def\bG{\mathbf G}              
\def\bI{\mathbf I}              
\def\bK{\mathbf K}              \def\bL{\mathbf L}
              
              \def\bP{\mathbf P}

              \def\bX{\mathbf X}

\def\ip#1{\left<#1\right>}

\def \Tsem      {\mbox{\sf\bf t\kern-1pt t}}
\def \Fsem      {\mbox{\sf\bf f\kern-1pt f}}

\newcommand{\trans}{\Rightarrow}

\newcommand{\syndef}{\mbox{\tt::=}}
\newcommand{\synalt}{\;\mbox{\tt\large$|$}\;}

\newcommand{\sem}[1]{[\![ #1 ]\!]}
\newcommand{\eval}[1]{\cE(#1)}

\newcommand{\tup}[1]{\langle #1 \rangle}

\newcommand{\false}{\mbox{\bf false}}
\newcommand{\true}{\mbox{\bf true}}

\newcommand{\skipS}{\mbox{\bf skip}}
\newcommand{\stopS}{\mbox{\bf stop}}
\newcommand{\getsS}[2]{\varE{#1} \gets #2}
\newcommand{\ranS}[2]{\varE{#1} ~{\tt ?\!=}~ #2}
\newcommand{\ifS}[3]{\mbox{\bf if}~#1~\mbox{\bf then}~#2~\mbox{\bf else}~#3}
\newcommand{\whileS}[2]{\mbox{\bf while}~#1~\mbox{\bf do}~#2}

\newcommand{\prg}[1]{{\mbox{\tt #1}}}

\newtheorem{definition}{Definition}

\newtheorem{example}{Example}
\newcommand{\AnalysisA}{\hbox{\sl Analysis}_{\circ}}
\newcommand{\AnalysisB}{\hbox{\sl Analysis}_{\bullet}}
\newcommand{\AnalysisC}{{\sf Analysis}}

\newcommand{\Lub}{\bigsqcup}

\newcommand{\init}{\mbox{\sl init}}
\newcommand{\final}{\mbox{\sl final}}

\newcommand{\flow}{\hbox{\sl flow}}

\newcommand{\FV}{\hbox{\sl FV}}

\newcommand{\States}{\mbox{\bf State}}
\newcommand{\Confs}{\mbox{\bf Conf}}

\newcommand{\Blocks}{\mbox{\bf Block}}
\newcommand{\Stmts}{\mbox{\bf Stmt}}
\newcommand{\Dists}{\mbox{\bf Dist}}

\newcommand{\Values}{\mbox{\bf Value}}

\newcommand{\Vars}{\mbox{\bf Var}}

\newcommand{\Labs}{\mbox{\bf Lab}}

\newcommand{\RD}{\hbox{\sl RD}}

\newcommand{\LV}{\hbox{\sl LV}}
\newcommand{\killLV}{\hbox{\sl kill}_{\sf LV}}
\newcommand{\genLV}{\hbox{\sl gen}_{\sf LV}}
\newcommand{\LVentry}{{\sf LV}_{\hbox{\scriptsize{\it entry}}}}
\newcommand{\LVexit}{{\sf LV}_{\hbox{\scriptsize{\it exit}}}}
\newcommand{\skipL}[1]{[\mbox{\tt skip}]\pp{#1}}

\newcommand{\getsL}[3]{[#1 ~\mbox{\tt :=}~ #2]\pp{#3}}
\newcommand{\ranL}[3]{[#1 ~\mbox{\tt ?=}~ #2]\pp{#3}}

\newcommand{\whileL}[3]{\mbox{\tt while}~[#1]\pp{#3}~\mbox{\tt do}~#2~\mbox{\tt od}}

\newcommand{\ifL}[4]{\mbox{\tt if}~[#1]\pp{#4}~\mbox{\tt then}~#2~\mbox{\tt else}~#3~\mbox{\tt fi}}

\newcommand{\pp}[1]{\mbox{${}^{#1}$}}
\renewcommand{\skipS}{\mbox{\tt skip}}
\renewcommand{\stopS}{\mbox{\tt stop}}
\renewcommand{\getsS}[2]{#1 ~\mbox{\tt :=}~ #2}
\renewcommand{\ranS}[2]{#1 ~\mbox{\tt ?=}~ #2}

\renewcommand{\whileS}[2]{\mbox{\tt while}~#1~\mbox{\tt do}~#2~\mbox{\tt od}}
\renewcommand{\ifS}[3]{\mbox{\tt if}~#1~\mbox{\tt then}~#2~\mbox{\tt else}~#3~\mbox{\tt fi}}
\newcommand{\Rstop}{ 
$\tup{\stopS,s} 
 {\trans}_{1} 
 \tup{\stopS,s}
$}
\newcommand{\Rskip}{ 
$\tup{\skipS,s} 
 {\trans}{}_{1} 
 \tup{\stopS,s}
$}
\newcommand{\Rassign}{ 
$\tup{\getsS{v}{e},s} 
 {\trans}_{1} 
 \tup{\stopS,s[v \mapsto \eval{e}s]}
$} 
\newcommand{\Rrandom}{ 
$\tup{\ranS{v}{\rho},s} 
 {\trans}_{\rho(r)}
 \tup{\stopS,s[v \mapsto r]}
$} 
\newcommand{\RseqA}{ 
$\frac{
  \displaystyle\tup{S_1,s}       {\trans}{}_{p} \tup{S'_1,s'}
}{%
  \displaystyle\tup{S_1 ; S_2,s} {\trans}{}_{p} \tup{S'_1 ; S_2,s'}
}
$} 
\newcommand{\RseqB}{
$\frac{%
  \displaystyle\tup{S_1,s}       {\trans}{}_{p} \tup{\stopS,s'}%
}{%
  \displaystyle\tup{S_1 ; S_2,s} {\trans}{}_{p}  \tup{S_2,s'}
}
$} 
\newcommand{\Riftrue}{ 
$\tup{\ifS{b}{S_1}{S_2},s}
 {\trans}{}_{1}
 \tup{S_1,s}
$} 
\newcommand{\Riffalse}{ 
$\tup{\ifS{b}{S_1}{S_2},s}
 {\trans}{}_{1}
 \tup{S_2,s}
$} 
\newcommand{\Rwhiletrue}{ 
$\tup{\whileS{b}{S},s}
 {\trans}{}_{1}
 \tup{S;~\whileS{b}{S},s}
$}
\newcommand{\Rwhilefalse}{ 
$\tup{\whileS{b}{S},s}
 {\trans}{}_{1}
 \tup{\stopS,s}
$} 

\title{Probabilistic data flow analysis: a linear equational approach}
\author{Alessandra Di Pierro
\institute{
University of Verona\\
Verona, Italy}
\email{alessandra.dipierro@univr.it}
\and
Herbert Wiklicky
\institute{Imperial College London\\
London, UK}
\email{\quad herbert@doc.ic.ac.uk}
}
\def\titlerunning{Equational Data Flow}
\def\authorrunning{A. Di Pierro, H. Wiklicky}
\begin{document}
\maketitle

\begin{abstract}
Speculative optimisation relies on the estimation of the probabilities that certain 
properties of the control flow are fulfilled. Concrete or estimated branch probabilities 
can be used for searching and constructing advantageous speculative and bookkeeping 
transformations.
We present a probabilistic extension of the classical equational approach to data-flow 
analysis that can be used to this purpose. More precisely, we show how the probabilistic 
information introduced in a control flow graph by branch prediction can be used to 
extract a system of linear equations from a program and present a method for calculating 
correct (numerical) solutions.
\end{abstract}


\section{Introduction}
\label{Introduction}

In the last two decades probabilistic aspects of 
software have become a particularly popular subject of research. The reason for
this is arguably in  {\em economical} and {\em resource conscious} questions
involving modern computer systems. While program verification and
analysis originally focused on qualitative issues, e.g. whether code 
is correct or if compiler optimisations are valid, the focus is now more often also
on the costs of  operations.

Speculative optimisation is part of this trend; it plays an important role in the design of 
modern compiler and run time architectures. 
A speculative approach has been adopted in various models 
where cost optimisation claims for a more optimistic interpretation of the results of a
program analysis. It is in fact often the case that  possible
optimisations are discarded because the analysis cannot guarantee their correctness. 
The alternative to this sometimes overly pessimistic analysis is to speculatively
assume in those cases that optimisations are correct and then eventually backtrack
and redo the computation if at a later check the assumption turns out to be incorrect.


Speculative optimisation relies on the optimal estimation 
of the probabilities that certain properties of the control flow are fulfilled. This is
different from the classical (pessimistic) thinking where one aims in 
providing bounds for what can happen during execution \cite{Festschrift06}.

A number of frameworks and tools to analyse systems's probabilistic aspects 
have been developed, which can be seen as probabilistic versions of classical
techniques such as model checking and abstract interpretation. To provide a basis 
for such analysis various semantical model involving discrete and continuous 
time and also non-deterministic aspects have been developed (e.g. DTMCs, 
CTMCs, MDPs, process algebraic approaches etc.). There also exist  
some powerful tools which implement these methods, e.g. 
PRISM \cite{PRISM04},  just to name one. 

Our own contribution in this area has been a probabilistic version of the abstract
interpretation framework \cite{CousotCousot77a}, called
 Probabilistic Abstract Interpretation (PAI) \cite{PPDP00,APLAS07}. 
This analysis framework, in its basic form, is concerned with purely probabilistic, 
discrete time models. Its purpose is to give optimal estimates of the probability 
that a certain property holds rather than providing probabilities bounds. 
As such, we think it is well suited as a base for speculative optimisation.

The aim of this paper is to provide a framework for a probabilistic analysis of
programs in the style of a classical data flow approach \cite{NielsonEtAl99,Dragon2}. 
In particular, we are interested in a formal basis for (non-static) branch 
prediction. The analysis technique we present consists of three phases: (i) abstract branch 
prediction, (ii) specification of the actual data-flow equations based on the 
estimates of the branch probabilities, and (iii) finding solutions. We will use 
vector space structures to specify the properties and analysis of a program. 
This allows for the construction of solutions via numerical (linear algebraic) 
methods as opposed to the lattice-theoretic fixed-point construction of the classical analysis.





\section{A Probabilistic Language}
\label{Language}

\subsection{Syntax and Operational Semantics}

We use as a reference language a simple imperative language whose  
syntax is given in Table~\ref{Syntax}.
 \begin{table}[t]
 \begin{center}
 \begin{minipage}{4cm}
 \begin{center}
 \begin{tabular}{rcl}
 $S$ & $\syndef$ & $\skipS$ \\
     & $\synalt$ & $\getsS{x}{e(x_1,\ldots,x_n)}$ \\
     & $\synalt$ & $\ranS{x}{\rho}$ \\
     & $\synalt$ & $S_1$\verb|;| $S_2$ \\ 
     & $\synalt$ & $\ifS{b}{S_1}{S_2}$ \\
     & $\synalt$ & $\whileS{b}{S}$ \\
 \end{tabular}
 \end{center}
 \end{minipage}
 \hspace{2cm}%
 \begin{minipage}{6cm}
 \begin{center}
 \begin{tabular}{rcl}
 $S$ & $\syndef$ & $\skipL{\ell}$ \\
     & $\synalt$ & $\getsL{x}{e(x_1,\ldots,x_n)}{\ell}$\\
     & $\synalt$ & $\ranL{x}{\rho}{\ell}$ \\
     & $\synalt$ & $S_1$\verb|;| $S_2$ \\ 
     & $\synalt$ & $\ifL{b}{S_1}{S_2}{\ell}$ \\
     & $\synalt$ & $\whileL{b}{S}{\ell}$ \\
 \end{tabular}
 \end{center}
 \end{minipage}
 \end{center}
 \caption{The syntax}
 \label{Syntax}
 \end{table}
Following the approach in \cite{NielsonEtAl99}\ we extend this syntax with unique 
program labels $\ell\in\Labs$ in order to be able to refer to certain program 
points during the analysis.

The dummy statement $\skipS$ has no computational effect. For
the arithmetic {\em expressions} $e(x_1,\ldots,x_n)$ on the right hand 
side (RHS) of the assignment as well as for the tests $b=b(x_1,\ldots,x_n)$ in {\tt if}
and {\tt while} statements, we leave the details of the syntax open as 
they are irrelevant for our treatment.
The RHS of a random assignment $\ranS{x}{\rho}$ is a distribution $\rho$ over some
set of values with the meaning that $x$ is assigned one of the possible constant values $c$
with probability $\rho(c)$. 



An operational semantics in the SOS style is given in 
Table~\ref{SOSTable}. This defines a probabilistic transition relation on
configurations in $\Confs=\Stmts\times\States$ with $\Stmts$ the set of 
all statements in our language together with $\stopS$ which indicates termination
and $\States=\Vars\rightarrow\Values$.
The details of the semantics of arithmetic and boolean expressions
$\sem{a}=\cE(a)$ and $\sem{b}=\cE(b)$ respectively are again left open in 
our treatment here and can be found in \cite{Bertinoro10}.

\newcommand{\putSOS}{
\begin{table}[t]
\begin{center}
\begin{tabular}{ll}
\begin{tabular}{lll}
{\bf R0}       & \Rstop      & \\
{\bf R1}       & \Rskip      & \\
{\bf R2}       & \Rassign    & \\
{\bf R3}       & \Rrandom    & 
\end{tabular}
&
\begin{tabular}{lll}
{\bf R4${}_1$} & \RseqA      & \\[1em]
{\bf R4${}_2$} & \RseqB      & 
\end{tabular}
\\
\multicolumn{2}{l}{
\begin{tabular}{lll}
{\bf R5${}_1$} & \Riftrue    & if $\eval{b}s=\true$ \\
{\bf R5${}_2$} & \Riffalse   & if $\eval{b}s=\false$\\
{\bf R6${}_1$} & \Rwhiletrue & if $\eval{b}s=\true$ \\
{\bf R6${}_2$} & \Rwhilefalse& if $\eval{b}s=\false$
\end{tabular}
}
\end{tabular}
\end{center}
\label{SOSTable}
\caption{The rules of the SOS semantics}
\end{table}
}


\subsection{Computational States} \putSOS

In any concrete computation or execution -- even when it is involving 
probabilistic elements -- the computational situation is uniquely defined 
by a mapping $s: \Vars\rightarrow\Values$ to which we refer to as a
{\em classical state}. Every variable in $\Vars$ has a unique value 
in $\Values$ possibly including $\bot\in\Values$ to indicate undefinedness.
We denote by $\States$ the set of all classical states.

In order to keep the mathematical treatment simple we will assume here 
that every variable can take values in a finite set $\Values$.  These sets 
can be nevertheless quite large and cover, for example, all finitely 
representable integers on a given machine.

For a finite set $X$ we denote by $\cP(X)$ the power-set of $X$ and by  $\cV(X)$ 
the free vector space over $X$, i.e. the set of formal linear combinations of
elements in $X$. We represent vectors via their coordinates $(x_1,\ldots,x_n)$ 
as rows, i.e. elements in $\R^{|X|}$ with $|X|$ denoting the cardinality of $X$
and use post-multiplication with matrices representing linear maps, i.e. 
$\bA(x)=x\cdot\bA$.
The set $\Dists(X)$ of distributions on $X$ -- i.e. $\rho: X \rightarrow [0,1]$ 
and $\sum_i \rho(x_i) = 1$ -- clearly correspond to a 
sub-set of $\cV(X)$.
We will also use a tuple notation for distributions: $\rho= \{ \ip{a,\frac{1}{2}}, \ip{b,\frac{1}{4}}, 
\ip{c,\frac{1}{4} }\}$ will denote a distribution where $a$ has probability 
$\rho(a)=\frac{1}{2}$ and $b$ and $c$ both have probability $\frac{1}{4}$. 
For uniform distributions we will simply specify the underlying set, 
e.g. $\{a,b,c\}$ instead of $\ip{a,\frac{1}{3}}, \ip{b,\frac{1}{3}}, 
\ip{c,\frac{1}{3} }\}$.

The tensor product is an essential element of the description of  probabilistic states. 
The tensor product\footnote{More precisely, the Kronecker 
product -- the coordinate based version of the abstract concept of a tensor product.}\
of two vectors $(x_1,\ldots,x_n)$ and $(y_1,\ldots,y_m)$ is given by  
$(x_1y_1,\ldots,x_1y_m,\ldots,x_ny_1,\ldots,x_ny_m)$ an $nm$ dimensional 
vector. Similarily for matrices. The tensor product of two vector spaces 
$\cV\otimes\cW$ can be defined as the formal linear combinations of the tensor 
products $v_i \otimes w_j$ with $v_i$ and $w_j$ base vectors in $\cV$ and 
$\cW$, respectively. For further details we refer e.g to \cite[Chap.~14]{Roman05}.

Importantly, the isomorphism $\cV(X \times Y) = \cV(X) \otimes \cV(Y)$
allows us to identify set of all distributions on the cartesian product of two 
sets with the tensor product of the spaces of distributions on $X$ and $Y$. 

We define a {\em probabilistic state} $\si$ as any probability distribution 
over classical states, i.e. $\si\in\Dists(\States)$. This can also be seen as 
$\si\in\cV(\States) =
\cV(\Vars\rightarrow\Values) = 
\cV(\Values^{|\Vars|}) =
\cV(\Values)^{\otimes v}$
the $v$-vold tensor product of $\cV(\Values)$ with $v=|\Vars|$.

\label{linrep}
In our setting we represent (semantical) functions and predicates or tests as
linear operators on the probabilistic state space, i.e. as matrices. For any 
function $f:X \mapsto Y$ we define a linear representation $|X|\times|Y|$ matrix by:
\[
(\bF_f)_{ij}
=
(\bF(f))_{ij}
=
(\bF)_{ij}
=
\left\{
\begin{array}{cl}
1 & \mbox{if}~ f(x_i) = y_j \\
0 & \mbox{otherwise}.
\end{array}
\right.
\]
where we assume some fixed enumeration on both $X$ and $Y$.
For an equivalence relation on $X$ we can also represent the function which 
maps every element in $X$ to its equivalence class $c:x\mapsto[x]$ in this 
way. Such a {\em classification matrix} contains in every row exactly one 
non-zero entry $1$. Classification matrices  (modulo reordering of indices)  
are in a one-to-one correspondence with the equivalence relations on a set $X$ and 
we will use them to define probabilistic abstractions for our analysis (cf. Section~\ref{ap}).
A predicate $p:X\rightarrow\{\true,\false\}$ is represented by a diagonal 
$|X|\times|X|$ matrix:
\[
(\bP_p)_{ij}
=
(\bP(p))_{ij}
=
(\bP)_{ij}
=
\left\{
\begin{array}{cl}
1 & \mbox{if}~ i=j ~\mbox{and}~ p(x_i) = \true \\
0 & \mbox{otherwise}.
\end{array}
\right.
\]



\subsection{Probabilistic Abstraction}
\label{PAI}

The analysis technique we  present in this paper will make use of a particular  notion
of abstraction of the state space (given as a vector space) which is formalised in 
terms of Moore-Penrose pseudo-inverse \cite{Roman05}.

\begin{definition}
Let $\cC$ and $\cD$ be two finite dimensional vector spaces, and let 
$\bA: \cC \to \cD$ be a linear map between
them. The linear map $\bA^\dagger = \bG: \cD \to \cC$ is the
{\em Moore-Penrose pseudo-inverse} of $\bA$ iff
\[
\bA \circ \bG = \bP_{A} 
~\mbox{ and }~ 
\bG \circ \bA = \bP_{G}
\]
where $\bP_{A}$ and $\bP_{G}$ denote orthogonal projections onto the ranges 
of $\bA$ and $\bG$.
\end{definition}
An operator or matrix is an {\em orthogonal projection} if $\bP^* = \bP^2 = \bP$ 
where ${.}^*$ denotes the {\em adjoint} which for real matrices correspond
simply to the transpose matrix $\bP^*=\bP^t$ \cite[Ch~10]{Roman05}. 

For invertible matrices the Moore-Penrose pseudo-inverse is the same as the
inverse. 
A special example is the {\em forgetful abstraction} $\bA_f$ which corresponds 
to a map $f:X\rightarrow\{\ast\}$ which maps all elements of $X$ onto a single 
abstract one. It is represented by a $|X|\times1$ matrix containing only $1$, and its
Moore-Penrose pseudo-inverse is given by $1\times|X|$ matrix with
all entries $\frac{1}{|X|}$. 

The Moore-Penrose pseudo-inverse allows us to construct the closest, in a 
least square sense (see for example \cite{CampbellMeyer79,BenIsraelGreville03}),
approximation $\bF^\#: \cD \rightarrow \cD$ of a concrete linear operator 
$\bF: \cC \rightarrow \cC$ for a given abstraction $\bA:\cC\rightarrow\cD$ as
\[
\bF^\# = 
\bA^\dagger\cdot\bF\cdot\bA =
\bG\cdot\bF\cdot\bA =
\bA \circ \bF \circ \bG.
\]
This notion of probabilistic abstraction is central in the
Probabilistic Abstract Interpretation (PAI)  framework.
For further details we refer to e.g.  \cite{Bertinoro10}. 
As we will use this notion later for abstracting branching probabilities, it is
important here to point out the guarantees that such abstractions are able to provide.
In fact,  these are not related to any correctness notion in the classical sense.
The theory of the least-square approximation
\cite{Deutsch01,BenIsraelGreville03}\ tells us that if $\cC$ and $\cD$ be two
finite dimensional vector spaces, $\bA: \cC \mapsto \cD$ a linear map between
them, and $\bA^\dagger = \bG: \cD\mapsto \cC$ its Moore-Penrose
pseudo-inverse, then the vector $x_0= y \cdot\bG $ is the one minimising the
distance between $x\cdot \bA$, for any vector $x$ in $\cC$, and $y$, i.e.
\[
\inf_{x\in\cC} \|x\cdot \bA - y\| = \|x_0\cdot \bA  - y\|.
\]

This guarantees that our probabilistic abstractions  
correspond to the {\bf closest} approximations in a metric sense of the concrete situations, as
they are constructed using the Moore-Penrose pseudo-inverse. 
 

\section{Data-Flow Analysis}
\label{Analysis}


Data-flow analysis is based on a statically determined flow relation. This is defined
in terms of two auxiliary operations, namely $\init:\Stmts \rightarrow \Labs$ and 
$\final:\Stmts \rightarrow \cP(\Labs)$, defined as follows:


\begin{center}
\begin{minipage}{6cm}
\[
\begin{array}{l}
\init([\skipS]^{\ell})            = \ell \\
\init([\getsS{v}{e}]^{\ell})      = \ell \\
\init([\ranS{v}{e}]^{\ell})       = \ell \\
\init(S_1 ; S_2)                 = \init(S_1) \\
\init(\ifS{[b]^{\ell}}{S_1}{S_2})  = \ell \\
\init(\whileS{[b]^{\ell}}{S})     = \ell 
\end{array}
\]
\end{minipage}
\begin{minipage}{8cm}
\[
\begin{array}{l}
\final([\skipS]^{\ell})            = \{ \ell \} \\
\final([\getsS{v}{e}]^{\ell})      = \{ \ell \} \\
\final([\ranS{v}{e}]^{\ell})        = \{ \ell \} \\
\final(S_1 ; S_2)                 = \final(S_2) \\
\final(\ifS{[b]^{\ell}}{S_1}{S_2})  = \final(S_1) \cup \final(S_2) \\
\final(\whileS{[b]^{\ell}}{S})     = \{ \ell \}. 
\end{array}
\]
\end{minipage}
\end{center}

The control flow $\cF(S)$ in $S\in\Stmts$ is defined via the 
function $\flow: \Stmts \rightarrow \cP(\Labs \times \Labs)$:
\[
\begin{array}{c}
\flow([\skipS]^{\ell}) = 
\flow([\getsS{v}{e}]^{\ell}) =      
\flow([\ranS{v}{e}]^{\ell}) =  
\emptyset 
\\
\flow(S_1 ; S_2) =
\flow(S_1) \cup \flow(S_2) \cup 
\{ (\ell,\init(S_2)) \mid \ell \in \final(S_1) \} \\
\flow(\ifS{[b]^{\ell}}{S_1}{S_2}) = 
\flow(S_1) \cup \flow(S_2) \cup
\{ (\ell,\underline{\init(S_1)}), (\ell,\init(S_2)) \} \\
\flow(\whileS{[b]^{\ell}}{S}) = 
\flow(S) \cup \{ (\ell,\underline{\init(S)}) \}\cup
         \{ (\ell',\ell) \mid \ell'\in\final(S) \}
\end{array}
\]

The definition of flow only records that a certain control flow step is possible. For
tests $b$ in conditionals and loops we indicate the branch
corresponding to the case when the test is successful by underlining it. We 
identify a statement $S$ with the block $[S]^{\ell}$ that contains it and with the 
(unique) label $\ell$ associated to the block. We will denote by $\Blocks=\Blocks(P)$ 
the set of all the blocks occurring in $P$, and use indistinctly $\Blocks$ 
and $\Labs$ to refer to blocks.


 \subsection{Monotone Framework} 
 
The classical data-flow analysis is made up of two components:
a ``local'' part which describes how the information representing the analysis 
changes when execution passes through a given block/label, and a ``global'' 
collection part
which describes how information is accumulated when a number of different
control flow paths (executions) come together.

This is formalised in a general scheme, called Monotone Framework in 
\cite[Section~2.3]{NielsonEtAl99}, where a data-flow analysis is defined via a number 
of equations over the lattice $L$ modelling the property to be analysed. 
For every program label $\ell$ we have two equations: one describing 
the generalised `entry'  in terms of the generalised `exit' of the block in question, 
and the other describing `exit'  in terms of `entry' -- for forward analysis we 
have $\circ$=entry and  $\bullet$=exit, for a backward analysis the situation 
is reversed.
\begin{eqnarray*}
\AnalysisB(\ell) & = & f_{\ell}(\AnalysisA (\ell))  \\
\AnalysisA(\ell) & = & 
      \left\{
      \begin{array}{l}
      \iota, \hbox{if } \ell \in E\\
      \Lub \{ \AnalysisB (\ell') \mid (\ell',\ell) \in F\}, \hbox{otherwise}
      \end{array}\right.     
\end{eqnarray*}

For the typical classical analyses, such as Live Variable \LV\ and Reaching 
Definition \RD, the property lattice $L$ is often the power-set of some underlying 
set (like $\Vars$ as in the case of the LV analysis). 
For a may-analysis the collecting operation $\sqcup$ of $L$ is  represented 
by set union $\cup$ and for must-analysis it is the intersection operation $\cap$.
The flow relation $F$ can be the forward or backward flow. $\iota$ specifies the initial 
or final analysis information on ``extreme'' labels in $E$, where 
$E$ is $\{\init(S_\star)\}$ or $\final(S_\star)$,
and $f_\ell$ is the transfer function 
associated with $B^\ell \in \Blocks(S)$ \cite[Section~2.3]{NielsonEtAl99}.


\subsection{Live Variable Analysis}
\label{LiveVariable Analysis}

We will illustrate the basic principles of the equational approach
to data flow analysis by considering Live Variable analysis  (\LV) following the presentation
in \cite[Section~2.1]{NielsonEtAl99}. The problem is to identify at any program
point those variables which are {\em live}, i.e. which may later be used
in an assignment or test.

There are two phases of  classical \LV\ analysis: 
(i) formulation of data-flow equations as set equations (or more generally 
over a property lattice $L$), (ii) finding or constructing solutions to these equations, 
for example, via a fixed-point construction. 
%
%
%
In the classical analysis we associate to every program point or label $\ell$ --
to be precise the entry and the exit of each label --
the information which describes (a super-set of) those variables which are
alive at this program point. 

Based on the auxiliary functions $\genLV: \Blocks \rightarrow \cP(\Vars)$ 
and $\killLV: \Blocks \rightarrow \cP(\Vars)$ which only depend on the syntax 
of the local block $[B]^\ell$ and are defined as
\begin{center}
\begin{minipage}{6cm}
\[
\begin{array}{rcl}
\killLV([\getsS{x}{a}]^{\ell})    & = & \{ \prg{x} \} \\
\killLV([\ranS{x}{\rho}]^{\ell}) & = & \{ \prg{x} \} \\
\killLV([\skipS]^{\ell})              & = & \emptyset     \\
\killLV({[b]}^{\ell})                    & = & \emptyset
\end{array}
\]
\end{minipage}
\begin{minipage}{6cm}
\[
\begin{array}{rcl}
\genLV([\getsS{x}{a}]^{\ell})      & = & \FV(a)    \\
\genLV([\ranS{x}{\rho}]^{\ell})   & = & \emptyset \\
\genLV([\skipS]^{\ell})                & = & \emptyset \\
\genLV({[b]}^{\ell})                      & = & \FV(b)
\end{array}
\]
\end{minipage}
\end{center}
we can define the transfer functions for the \LV\ analysis
\(
f^{LV}_\ell:  \cP(\Vars_\star)  \rightarrow \cP(\Vars_\star)
\)\
by 
\[
f^{LV}_\ell(X) = X\setminus\killLV([B]^\ell) \cup \genLV([B]^\ell)
\] 

This allows us to define equations over the property space $L=\cP(\Vars)$,
i.e. set equations, which associate to every label entry and exit the
analysis information $\LVentry: \Labs \rightarrow \cP(\Vars)$ and
$\LVexit: \Labs \rightarrow \cP(\Vars)$.
These set equations are of the general form for a backward may 
analysis:
\begin{eqnarray*}
\LVentry(\ell) & = & f^{LV}_\ell(\LVexit(\ell)) \\
\LVexit(\ell)    & = & \bigcup_{(\ell,\ell')\in\flow} \LVentry(\ell')
\end{eqnarray*}
At the beginning of the analysis (i.e. for final labels, as this is a 
backward analysis) we set $\LVexit(\ell) = \emptyset$.


\begin{example}
\label{running}
Consider  the following program:
\[
\begin{array}{l}
\ranL{x}{\{0,1\}}{1};\;
\ranL{y}{\{0,1,2,3\}}{2};\;
\getsL{x}{x+y \bmod 4}{3};\; \\
\ifL{x>2}{
\getsL{z}{x}{5}
}{
\getsL{z}{y}{6}
}{4}
\end{array}
\]
Although the program is probabilistic we still can perform a classical
analysis by considering non-zero probabilities simply as possibilities.
The flow is given by
\(
\{ 
(1,2), (2,3), (3,4),
(4,\underline5), (4,6)
\}.
\)

With the auxiliary functions $\killLV$ and $\genLV$ we can now specify the data-flow equations:

\hspace*{-7mm}
\begin{minipage}{3.9cm}
\[
\begin{array}{|c|c|c|}
\hline
& \genLV(\ell) & \killLV(\ell) \\
\hline
1 & \emptyset & \{x\} \\
2 & \emptyset & \{y\} \\
3 & \{x,y\} & \{x\} \\
4 & \{x\} & \emptyset \\
5 & \{x\} & \{z\} \\
6 & \{y\} & \{z\} \\
\hline
\end{array}
\]
\end{minipage}
\begin{minipage}{6.5cm}
\vspace{3mm}
\begin{eqnarray*}
\LVentry(1) & = & \LVexit(1)\setminus\{x\}  \\
\LVentry(2) & = & \LVexit(2)\setminus\{y\}  \\
\LVentry(3) & = & \LVexit(3)\setminus\{x\} \cup\{x,y\} \\
\LVentry(4) & = & \LVexit(4) \cup \{x\} \\
\LVentry(5) & = & \LVexit(5)\setminus\{z\} \cup\{x\} \\
\LVentry(6) & = & \LVexit(6)\setminus\{z\} \cup\{y\} \\
\end{eqnarray*}
\end{minipage}
\begin{minipage}{6cm}
\vspace{3mm}
\begin{eqnarray*}
\LVexit(1) & = & \LVentry(2) \\
\LVexit(2) & = & \LVentry(3) \\
\LVexit(3) & = & \LVentry(4) \\
\LVexit(4) & = & \LVentry(5) \cup \LVentry(6) \\
\LVexit(5) & = & \emptyset \\
\LVexit(6) & = & \emptyset \\
\end{eqnarray*}
\end{minipage}

Then the classical \LV\ analysis of our program gives the solutions:

\begin{center}
\begin{minipage}{5cm}
\begin{eqnarray*}
\LVentry(1) & = & \emptyset \\
\LVentry(2) & = & \{x\} \\
\LVentry(3) & = & \{x,y\} \\
\LVentry(4) & = & \{x,y\} \\
\LVentry(5) & = & \{x\} \\
\LVentry(6) & = & \{y\} \\ 
\end{eqnarray*}
\end{minipage}
\begin{minipage}{5cm}
\begin{eqnarray*}
\LVexit(1) & = & \{x\} \\ 
\LVexit(2) & = & \{x,y\} \\
\LVexit(3) & = & \{x,y\} \\
\LVexit(4) & = & \{x,y\} \\
\LVexit(5) & = & \emptyset \\
\LVexit(6) & = & \emptyset. \\
\end{eqnarray*}
\end{minipage}
\end{center}

\end{example}

\section{The Probabilistic Setting}

In order to specify a probabilistic data flow analysis using the analogue of the 
classical equational approach (as presented in the previous sections), we 
have to define the main ingredients of the analysis in a probabilistic setting
namely a vector space as property space (replacing the property lattice $L$), 
a linear operator representing the  transfer functions $f_\ell$, and a method 
for the information collection (in place of the $\Lub$ operation of the classical 
monotone framework). Moreover, as we will work with probabilistic states,
the second point implies that the control-flow
graph will be labelled by some probability information. 

As a {\bf property space} we consider distributions $\Dists(L)\subseteq\cV(L)$ 
over a set $L$, e.g. the corresponding classical property space. For a relational analysis, 
where the classical property lattice corresponds to $L=L_1\times L_2$ (cf \cite{QAPL08}),
the probabilistic property space will be the tensor product
$\cV(L_1)\otimes\cV(L_2)$; this allows us to represent properties via joint probabilities 
which are able to express the dependency or correlation between states. 


We can define probabilistic {\bf transfer functions} by using the linear representation
of the classical $f_\ell$, i.e. a matrix $\bF_\ell=\bF_{f_\ell}$ as introduced above
in Section~\ref{linrep}.
%
In general, we will define a probabilistic transfer function by means of an
appropriate abstraction of the concrete semantics 
$\sem{[B]^\ell}$ of a given block $[B]^\ell$ according to PAI, i.e. 
$\bF_\ell = \bA^\dagger\sem{[B]^\ell}\bA$ for the relevant abstraction 
matrix $\bA$. 

In the classical analysis we treat tests $b$ non-deterministically, to avoid
problems with the potential undecidability of predicates. Moreover,
we take everything which is possible i.e. the collection of what can happen along the different execution 
paths, e.g. the two branches of an {\tt if} statement. In the probabilistic setting we 
{\bf collect information} by means of weighted sums, where the `weights' are 
the probabilities associated to each branch. These probabilities come 
from an estimation of the (concrete or abstract) branch probabilities and 
are propagated along the control flow graph representing the {\bf flow relation}.

\subsection{Control Flow Probabilities}
%
%

If we execute a program in classical states $s$ which have been chosen 
randomly according to some probability distribution $\rho$ then this
also induces a probability distribution on the possible control flow steps.

\begin{definition}
  Given a program $S_\ell$ with $\init(S_\ell)=\ell$ and a probability
  distribution $\rho$ on $\States$, the probability
  $p_{\ell,\ell'}(\rho)$ that the control is flowing from $\ell$ to $\ell'$ is 
  defined as:
  \[
  p_{\ell,\ell'}(\rho) = 
  \sum_{s}  
  \left\{ 
  p\cdot\rho(s) ~|~ \exists s'\mbox{ s.t. } \ip{S_\ell,s} \trans_p \ip{S_{\ell'},s'} 
  \right\}.
  \]
\end{definition}

In other words, if we provide with a certain probability $\rho(s)$ a
concrete execution environment or classical state $s$ for a program
$S_\ell$, then the control flow probability $p_{\ell,\ell'}(\rho)$ is the
probability that we end up with a configuration $\ip{S_{\ell'},\ldots}$ 
for whatever state in the successor configuration. 
%

\begin{example}
\label{1}
Consider the program:
\(
\ranL{x}{\{0,1\}}{1};\;
\ifL{x>0}{\skipL{3}}{\getsL{x}{0}{4}}{2}
\). 
We can have two possible states at label $2$, namely $s_0=[x\mapsto0]$
and  $s_1=[x\mapsto1]$. After the first statement has been executed 
in one of two possible ways (with any intial state $s$):
\begin{eqnarray*}
\lefteqn{
\ip{
\ranL{x}{\{0,1\}}{1};\;
\ifL{x>0}{\skipL{3}}{\getsL{x}{0}{4}}{2}, 
s}
\trans_{\frac{1}{2}} }\\
& \trans_{\frac{1}{2}} &
\ip{
\ifL{x>0}{\skipL{3}}{\getsL{x}{0}{4}}{2}, 
s_0}
\\or~~~
\lefteqn{
\ip{
\ranL{x}{\{0,1\}}{1};\;
\ifL{x>0}{\skipL{3}}{\getsL{x}{0}{4}}{2}, 
s}
\trans_{\frac{1}{2}} }\\
& \trans_{\frac{1}{2}} &
\ip{
\ifL{x>0}{\skipL{3}}{\getsL{x}{0}{4}}{2}, 
s_1}
\end{eqnarray*}
the distribution over states is obviously $\rho=\{
\ip{s_0,\frac{1}{2}} \ip{s_1,\frac{1}{2}} \}$. However, in each
execution path we have at any moment a definite value for $x$ (the
distribution $\rho$ describes a property of the set of all executions,
not of one execution alone).

The branch probability in this case (independently of the state $s$ and of any distribution $\rho$) 
is simply $p_{1,2}(\rho)=1$ because, although there are two possible execution
steps, the successor configurations are `coincidently' equipped with
the same program $\ifL{x>0}{\skipL{3}}{\getsL{x}{0}{4}}{2}$.

The successive control steps from label $2$ to $3$ and $4$, respectively,
both occur with probability $1$ as in each state $s_0$ and $s_1$ the value of $x$ is a definite one.
\begin{eqnarray*}
\ip{
\ifL{x>0}{\skipL{3}}{\getsL{x}{0}{4}}{2}, 
s_0}
& \trans_1 &
\ip{
\getsL{x}{0}{4},
s_0}
\\and~~~
\ip{
\ifL{x>0}{\skipL{3}}{\getsL{x}{0}{4}}{2}, 
s_1}
& \trans_1 &
\ip{
\skipL{3}, 
s_1}
\end{eqnarray*}

Thus the branch probabilities with $\rho=\{ \ip{s_0,\frac{1}{2}},
\ip{s_1,\frac{1}{2}} \}$ are $p_{2,3}(\rho)=\frac{1}{2}$ and
$p_{2,4}(\rho)=\frac{1}{2}$. In general for any $\rho=\{ \ip{s_0,p_0},
\ip{s_1,p_1} \}$ we have $p_{2,3}(\rho)=p_1$ and $p_{2,4}(\rho)=p_0$\
despite the fact that the transitions are deterministic. It is the
randomness in the probabilistic state that determines in this case the
branch probabilities.
\end{example}

For all  blocks in a control flow graph -- except  for the tests $b$ --
there is  always only one next statement $S_{\ell'}$ so that the branch
probability $p_{\ell,\ell'}(\rho)$ is always $1$ for all $\rho$. 
For tests $b$ in {\tt if} and {\tt while} 
statements we have only two different successor statements, one 
corresponding to the case where $[b]^\ell$ evaluates to $\true$ and 
one for $\false$. As the corresponding probabilities must sum up to 
$1$ we only need to specify the first case which we denote by 
$p_\ell(\rho)$.

The probability distributions over states at every execution 
point are thus critical for the analysis 
as they determine the branch 
probabilities for tests, and we need to provide them. The problem
is, of course that analysing these probabilities is nearly as
expensive as analysing the concrete computation or program
executions. It is therefore reasonable to investigate
abstract branch probabilities, based on classes of states, or abstract 
states. It is always possible to lift concrete distributions to ones
over (equivalence) classes.



\begin{definition}
  Given a probability distribution $\rho$ on \States\ and an
  equivalence relation $\sim$ on states then we denote by
  $\rho^\#=\rho^\#_\sim$ the probability distribution on the set of
  equivalence classes $\States^\#=\States/\!\!\sim$ defined by
  \[
  \rho^\# ([s]_\sim) = \sum_{s'\in[s]_\sim} \rho(s')
  \]
  where $[s]_\sim$ denotes the equivalence classes of  $s$ wrt $\sim$.
\end{definition}

%

\subsection{Estimating Abstract Branch Probabilities}
\label{ap}

In order to determine concrete or abstract branch probabilities we
need to investigate -- as we have seen in Example~\ref{1} -- the
interplay between distribution over states and the test $[b]^\ell$ we are
interested in. We need for this the linear representation $\bP_b$ of 
the test predicate $b$ as defined in Section~\ref{linrep}, which
%
%
for a given distribution over states determines
 a sub-distribution of those states that lead into one of the two branches
by filtering out those states where this happens. 

\begin{example}
Consider the simple program
\(
\ifL{x>=1}{\getsL{x}{x-1}{2}}{\skipL{3}}{1}
\) 
and assume that $x$ has values in $\{0,1,2\}$ (enumerated in the
obvious way). 
Then the test $b=(x>=1)$ is represented by the projection matrix:
\[
\bP(x>=1)=
\left(
\begin{array}{ccc}
0 & 0 & 0 \\
0 & 1 & 0 \\
0 & 0 & 1 \\
\end{array}
\right)
~\mbox{and}~~
\bP(x>=1)^\perp=
\left(
\begin{array}{ccc}
1 & 0 & 0 \\
0 & 0 & 0 \\
0 & 0 & 0 \\
\end{array}
\right)
=\bP(x=0)
\]
For any given concrete probability distribution over states
$\rho=\{ \ip{0,p_0}, \ip{1,p_1}, \ip{2,p_2} \} = (p_0,p_1,p_2)$
we can easily compute the probabilities to go from label $1$
to label $2$ as $\rho\bP(x>=1)=(0,p_1,p_2)$ and thus 
\[
p_{1,2}(\rho)=\|\rho\cdot\bP(x>=1)\|_1 = p_1+p_2,
\]
where $\|.\|_1$ is the 1-norm of vectors, i.e. $\|(x_i)_i\|=\sum_i |x_i|$, 
which we use here to aggregate the total probabilities. Similarly, for the
else branch, with $\bP^\perp=\bI-\bP$:
\[
p_{1,3}(\rho)=\|\rho\cdot\bP^\perp(x>=1)\|_1 = p_0.
\]
\end{example}

In general, the branching behaviour at a test $b$ is described by the projection operator 
$\bP(b)$ and its complement $\bP^\perp(b)=\bP(\neg b)$. 
  For a branching point $[b]^\ell$ with $(\ell,\underline\ell'),(\ell,\ell'')\in\flow$, we denote
 $\bP(b)$ by  $\bP(\ell,\ell')$ and $\bP(\neg b)=\bP(b)^\perp$ by $\bP(\ell,\ell'')$. 
  Each branch probability can be computed for any given input distribution
  as   $p_{\ell,\ell'}(\rho)=  \|\rho\bP(\ell,\ell')\|_1$ and $p_{\ell,\ell''}(\rho)= \|\rho\bP(\ell,\ell'')\|_1$,
  respectively.

Sometimes it could be useful or practically more appropriate to consider
abstract branch probabilities. These can be obtained by means of abstractions
on the state space corresponding to classifications $c:\States\rightarrow\States^\#$
that, as explained in Section~\ref{linrep}, can be lifted to {\em classification matrices}.
Given an equivalence relation $\sim$ on the states and its matrix representation 
$\bA_\sim$, we can compute the individual chance of abstract states (i.e.
equivalence classes of states) to take the \true\ or \false\ branch of a
test by multiplying the abstract distribution
$\rho^\#$ by an abstract version $\bP(b)^\#$ of $\bP(b)$ that
we can use to select those classes of states satisfying $b$. 
In doing so we must guarantee that:
\begin{eqnarray*}
\rho \bP(b) \bA & = &  \rho^\# \bP^\#(b) \\
\rho \bP(b) \bA & = & \rho \bA \bP^\#(b) \\
\bP(b) \bA & = & \bA \bP^\#(b)
\end{eqnarray*}
In order to give an explicit description of $\bP^\#$ we only would need
to multiply the last equation from the left with $\bA^{-1}$.
However, $\bA$ is in general not a square matrix and thus
not invertible. So we use instead the Moore-Penrose 
pseudo-inverse to have the closest, least-square approximation possible.
\begin{eqnarray*}
\bA^\dagger \bP(b) \bA & = & \bA^\dagger \bA \bP^\#(b) \\
\bA^\dagger \bP(b) \bA & = & \bP^\#(b) \\
\end{eqnarray*}

The abstract test matrix $\bP^\#(b)$ contains all the information
we need in order to estimate the abstract branch probabilities.
Again, we denote by $\bP(\ell,\ell')^\#=\bP^\#(b)$ and $\bP(\ell,\ell'')^\#=
\bP^\#(\neg b)=\bP^\#(b)^\perp$ for a branching point $[b]^\ell$ with 
$(\ell,\underline\ell'),(\ell,\ell'')\in\flow$. 


Branch prediction/predictors in hardware design has long history \cite{McFarling93,StylesLuk04}.
It is used at test points $[b]^\ell$ to allow pre-fetching of instructions of the expected branch 
before the test is actually evaluated. If the prediction is wrong the prefetched instructions 
need to be discarded and the correct ones to be fetched. Ultimately, wrong
predictions ``just'' lead to longer running times, the correctness of the program
is not concerned. It can be seen as a form of speculative optimisation.
Typical applications or cases where branch prediction is relevant is for nested
tests (loops or ifs). Here we get exactly the interplay between different tests and/or
abstractions. We illustrate this in the following example.

\begin{example}
Consider the following program that  counts the prime numbers.
\[
\getsL{i}{2}{1};\;
\whileL{i<100}{
  \ifL{prime(i)}{
	\getsL{p}{p+1}{4}
  }{
	\skipL{5}
  }{3};\ \getsL{i}{i+1}{6}
}{2}
\]

Within our framework we can simulate to a certain degree a history dependent 
branch prediction. If the variable $p$ has been updated in the previous iteration
it is highly unlikely it will so again in the next -- in fact that only happens in the
first two iterations. One can also interpret this as follows: For $i$ even the
branch probability $p_{3,4}(\rho_e)$ at label $3$ is practically zero for 
any reasonable distribution, e.g. a uniform distribution $\rho_e$, on evens.
To see this, we need to investigate only the form of 
\[
 \bP(prime(i))^\# = \bA_e^\dagger \bP(prime(i)) \bA_e,
\]
where $A_e$ is the abstraction corresponding to the classification in $even$ and $odd$.
\end{example}

In order to understand how an abstract property interacts with the branching in the program, 
as in the previous example we look at $\bA^\dagger \bP(b) \bA$ in order to 
evaluate how good a branch prediction is for a certain predicate/test $b$
if it is based on a certain abstraction/property $\bA$.
This is explained in the following example where we consider two 
properties/abstractions and corresponding tests.

\begin{example}
Let us consider two tests for numbers in the range $i=0,1,2,3,\ldots,n$):
\begin{center}
%
\begin{minipage}{7cm}
\[
\bP_e=
(\bP(\mbox{even}(n)))_{ii}=
\left\{
\begin{array}{cl}
1 & \mbox{if}~i=2k \\
0 & otherwise
\end{array}
\right.
\]
\end{minipage}
\begin{minipage}{7cm}
\[
\bP_p=
(\bP(\mbox{prime}(n)))_{ii}=
\left\{
\begin{array}{cl}
1 & \mbox{if}~\mbox{prime}(i) \\
0 & otherwise
\end{array}
\right.
\]
\end{minipage}
\end{center}
Likewise we can consider two corresponding abstractions
($j\in\{1=\true,2=\false\}$):
\begin{center}
%
\begin{minipage}{6cm}
\[
(\bA_e)_{ij}=
\left\{
\begin{array}{cl}
1 & \mbox{if}~i=2k+1~\wedge~j=2 \\
1 & \mbox{if}~i=2k~\wedge~j=1 \\
0 & otherwise
\end{array}
\right.
\]
\end{minipage}
\begin{minipage}{6cm}
\[
(\bA_p)_{ij}=
\left\{
\begin{array}{cl}
1 & \mbox{if}~\mbox{prime}(i)~\wedge~j=2 \\
1 & \mbox{if}~\neg\mbox{prime}(i)~\wedge~j=1 \\
0 & otherwise
\end{array}
\right.
\]
\end{minipage}
\end{center}

Then we can use $\bP^\#$ and its orthogonal complement,
$(\bP^\#)^\perp=\bI-\bP^\#$ to determine information about the quality of
a certain property or its corresponding abstraction via the number of false positives.
In fact, this will tell us how precise the  abstraction is with respect to tests 
(such as those controlling a loop or conditional). 
With rounding the values to 2 significant digits we get, for example the 
following results for different concrete ranges of the concrete values
$0,\ldots,n$.
\[
\begin{array}{rcccc}
&
\bA_e^\dagger\bP_p \bA_e &
\bA_e^\dagger\bP_p^\perp \bA_e &
\bA_p^\dagger\bP_e \bA_p &
\bA_p^\dagger\bP_e^\perp \bA_p 
\\[1em]
n=10 &
\left(
\begin{array}{rr}
0.20 & 0.00 \\ 
0.00 & 0.60 \\ 
\end{array}
\right)
&
\left(
\begin{array}{rr}
0.80 & 0.00 \\ 
0.00 & 0.40 \\ 
\end{array}
\right)
&
\left(
\begin{array}{rr}
0.25 & 0.00 \\ 
0.00 & 0.67 \\ 
\end{array}
\right)
&
\left(
\begin{array}{rr}
0.75 & 0.00 \\ 
0.00 & 0.33 \\ 
\end{array}
\right)
\\[1em]
n=100 & 
\left(
\begin{array}{rr}
0.02 & 0.00 \\ 
0.00 & 0.48 \\ 
\end{array}
\right)
&
\left(
\begin{array}{rr}
0.98 & 0.00 \\ 
0.00 & 0.52 \\ 
\end{array}
\right)
&
\left(
\begin{array}{rr}
0.04 & 0.00 \\ 
0.00 & 0.65 \\ 
\end{array}
\right)
&
\left(
\begin{array}{rr}
0.96 & 0.00 \\ 
0.00 & 0.35 \\ 
\end{array}
\right)
\\[1em]
n=1000 & 
\left(
\begin{array}{rr}
0.00 & 0.00 \\ 
0.00 & 0.33 \\ 
\end{array}
\right)
&
\left(
\begin{array}{rr}
1.00 & 0.00 \\ 
0.00 & 0.67 \\ 
\end{array}
\right)
&
\left(
\begin{array}{rr}
0.01 & 0.00 \\ 
0.00 & 0.60 \\ 
\end{array}
\right)
&
\left(
\begin{array}{rr}
0.99 & 0.00 \\ 
0.00 & 0.40 \\ 
\end{array}
\right)
\\[1em]
n=10000 & 
\left(
\begin{array}{rr}
0.00 & 0.00 \\ 
0.00 & 0.25 \\ 
\end{array}
\right)
&
\left(
\begin{array}{rr}
1.00 & 0.00 \\ 
0.00 & 0.75 \\ 
\end{array}
\right)
&
\left(
\begin{array}{rr}
0.00 & 0.00 \\ 
0.00 & 0.57 \\ 
\end{array}
\right)
&
\left(
\begin{array}{rr}
1.00 & 0.00 \\ 
0.00 & 0.43 \\ 
\end{array}
\right)
\\
\end{array}
\]
Note that  the positive and negative versions of these matrices always add up to the identity matrix $\bI$. 
Also, the entries in the upper left corner of $\bA_e^\dagger\bP_p \bA_e$ give us
information about the chances that an even number is also a prime number: For
small $n$ the percentage is a fifth (indeed $2$ is a prime and it is one out
of $5$ even numbers under $10$); the larger $n$ gets the less relevant is this
single even prime. With $\bA_p^\dagger\bP_e \bA_p$ we get the
opposite information: Among the prime numbers $\{2,3,5,7\}$ smaller than $10$
there is one which is even, i.e. 25\%; again this effect diminishes for larger
$n$. Finally, the lower right entry in these matrices gives us the percentage
that a non-prime number is odd and/or that an odd number is not prime,
respectively.
\end{example}


\subsection{Linear Equations Framework}

A general framework for our probabilistic data-flow analysis can be defined
in analogy with the classical monotone framework by defining the following
 linear equations: 
\begin{eqnarray*}
\AnalysisB(\ell) & = & \AnalysisA (\ell) \cdot \bF_\ell  \\
\AnalysisA(\ell) & = & 
      \left\{
      \begin{array}{l}
      \iota, \hbox{if } \ell \in E\\
      \sum \{ 
      	\AnalysisB (\ell') \cdot \bP(\ell',\ell)^\#
	\mid (\ell',\ell) \in F\}, \hbox{otherwise}
      \end{array}\right.     
 \end{eqnarray*}
 
 The first equation is a straight forward generalisation of the classical case, while the 
 second one is defined by means of the linear sums over vectors. 
 A simpler version is obtained by considering static branch prediction:
 \[
 \AnalysisA(\ell) =  \sum \{ 
      	p_{\ell',\ell} \cdot \AnalysisB (\ell') 
	\mid (\ell',\ell) \in F \}
 \]
 with $p_{\ell',\ell}$ is a numerical value representing a {\em static} branch probability. 
 
We have as many variables in this systems of equations as there are individual
equations. As a result we get unique solutions rather than least fix-points 
as in the classical setting.

\newcommand{\ProbA}{\hbox{\sl Prob}_{\circ}}
\newcommand{\ProbB}{\hbox{\sl Prob}_{\bullet}}
\newcommand{\ProbC}{{\sf Prob}}

This general scheme must be extended to include a preliminary phase 
of probability estimation if one wants to 
improve the quality of the branch prediction. In this case, 
the abstract state should carry two kinds of information: 
One, \ProbC, to provide estimates for probabilities, 
the other, \AnalysisC, to analyse the actual property in question. 
The same abstract branch probabilities $\bP(\ell',\ell)^\#$ -- which we obtain via 
\ProbC\ -- can then be used in both cases, but we have different information or
properties and different transfer functions for \ProbC\ and \AnalysisC.

\subsection{Probabilistic Live Variable Analysis}

We can use the previously defined probabilistic setting for a data flow analysis, 
to define a probabilistic version of the Live Variable analysis extending the 
one in \cite{NielsonEtAl99}\ in order to also cover for random 
assignments and to provide estimates for `live' probabilities.

The transfer functions, which describe how the program analysis information
changes when we pass through a block  $[B]^\ell$, is for the classical
analysis given via the two auxiliary functions $\genLV$ and $\killLV$ 
(cf. Example~\ref{running}).
Probabilistic versions of these operations can be defined as follows.
Consider two properties $d$ for `dead', and $l$ for `live' and the
space $\cV(\{0,1\})=\cV(\{d,l\})=\R^2$ as the property space corresponding to
a single variable. On this space define the operators:
\[
\bL=
\left(
\begin{array}{cc}
0 & 1 \\
0 & 1 \\
\end{array}
\right)
~~~\mbox{and}~~~
\bK=
\left(
\begin{array}{cc}
1 & 0 \\
1 & 0 \\
\end{array}
\right).
\]
The matrix $\bL$ changes the ``liveliness'' of a variable from whatever it
is (dead or alive) into alive, while $\bK$ does the opposite. The 
local transfer operators
\[
\bF_\ell=\bF^{LV}_\ell: \cV(\{0,1\})^{\otimes|\Vars|} \rightarrow \cV(\{0,1\})^{\otimes|\Vars|}
\]
for the block $\getsL{x}{a}{\ell}$ can thus be defined as (with $\bI$ the identity matrix)
\[
\bF_\ell = \bigotimes_{x_i\in\Vars} \bX_i
~~\mbox{with}~~
\bX_i =
\left\{
\begin{array}{cl}
\bL & \mbox{if}~x_i\in\FV(a) \\
\bK & \mbox{if}~x_i = x ~\wedge~ x_i\not\in\FV(a) \\
\bI   & \mbox{otherwise.} 
\end{array}
\right.
\]
and similarly for tests $[b]^\ell$
\[
\bF_\ell = \bigotimes_{x_i\in\Vars} \bX_i
~~\mbox{with}~~
\bX_i =
\left\{
\begin{array}{cl}
\bL & \mbox{if}~x_i\in\FV(b) \\
\bI   & \mbox{otherwise.} 
\end{array}
\right.
\]
For $\skipL{\ell}$ and random assignments $\ranL{x}{\rho}{\ell}$ we simply 
have $\bF_\ell = \bigotimes_{x_i\in\Vars} \bI$.

In the following example we demonstrate the use of our general framework for probabilistic 
data-flow analysis by defining a probabilistic \LV\ analysis for the program in Example~\ref{running}.

 \begin{example}
 \label{prunning}
 
For the program in Example~\ref{running}\ we present a \LV\ analysis based on
  concrete branch probabilities.  That means that in the first phase of the analysis 
  (which determines the branch probabilities) we will not abstract the values of $x$
  and $y$ (and ignore $z$ all together).
  If the concrete state of each variable is a value in $\{0,1,2,3\}$, then the
  probabilistic state is an element in $\cV(\{0,1,2,3\})^{\otimes3} = \R^{4^3}
  =\R^{64}$.
  The abstraction we use when we compute the concrete branch probabilities is
  $\bI\otimes\bI\otimes\bA_f$, i.e. $z$ is ignored. This allows us to
  reduce the dimensions of the probabilistic state space
  from $64$ down to just $16$. 
  The abstract transfer functions for the first $3$ statements
  are given in the Appendix. 
  
We can now compute the probability distribution at label $4$
for any given input distribution. The abstract 
transfer functions $\bF_5^\#$ and $\bF_6^\#$ are the identity 
as we have restricted ourselves only to the variables $x$ and 
$y$.

\newcommand{\Dentry}{{\sf Prob}_{\hbox{\scriptsize{\it entry}}}}
\newcommand{\Dexit}{{\sf Prob}_{\hbox{\scriptsize{\it exit}}}}

We can now set the linear equations for the joint distributions over $x$ and $y$ at
the entry and exit to each of the labels:

\begin{minipage}{7cm}
\begin{eqnarray*}
\Dentry(1) & = & \rho \\
\Dentry(2) & = & \Dexit(1) \\
\Dentry(3) & = & \Dexit(2) \\
\Dentry(4) & = & \Dexit(3) \\
\Dentry(5) & = & \Dexit(4)\cdot\bP_4^\# \\
\Dentry(6) & = & \Dexit(4)\cdot(\bI-\bP_4^\#) \\
\end{eqnarray*}
\end{minipage}
\begin{minipage}{7cm}
\begin{eqnarray*}
\Dexit(1) & = & \Dentry(1)\cdot\bF_1^\# \\
\Dexit(2) & = & \Dentry(1)\cdot\bF_2^\# \\
\Dexit(3) & = & \Dentry(1)\cdot\bF_3^\# \\
\Dexit(4) & = & \Dentry(4) \\
\Dexit(5) & = & \Dentry(5) \\ 
\Dexit(6) & = & \Dentry(6) \\ 
\end{eqnarray*}
\end{minipage}

These equations are easy to solve. In particular we can explicitly
determine 
\begin{eqnarray*}
\Dentry(5) & = & \rho\cdot\bF_1^\#\cdot\bF_2^\#\cdot\bF_3^\#\cdot\bP_4^\# \\
\Dentry(6) & = & \rho\cdot\bF_1^\#\cdot\bF_2^\#\cdot\bF_3^\#\cdot\bP_4^\#,
\end{eqnarray*}
that give us the static branch probabilities
$p_{4,5}(\rho)=\|\Dentry(5)\|_1=\frac{1}{4}$ and
$p_{4,6}(\rho)=\|\Dentry(6)\|_1=\frac{3}{4}$. 
These distributions can explicitly be computed
and do not depend on the initial distribution $\rho$.

We then perform a probabilistic \LV\ analysis using these probabilities 
as required. Using the abstract property space and the auxiliary operators
we get:

\begin{minipage}{8cm}
\begin{eqnarray*}
\LVentry(1) & = & \LVexit(1) \cdot (\bK \otimes \bI \otimes \bI) \\
\LVentry(2) & = & \LVexit(2) \cdot (\bI \otimes \bK \otimes \bI) \\ 
\LVentry(3) & = & \LVexit(3) \cdot (\bL \otimes \bL \otimes \bI)  \\
\LVentry(4) & = & \LVexit(4) \cdot (\bL \otimes \bI \otimes \bI) \\
\LVentry(5) & = & \LVexit(5) \cdot (\bL \otimes \bI \otimes \bK) \\
\LVentry(6) & = & \LVexit(6) \cdot (\bI \otimes \bL \otimes \bK) \\
\end{eqnarray*}
\end{minipage}
\begin{minipage}{6cm}
\begin{eqnarray*}
\LVexit(1) & = & \LVentry(2) \\
\LVexit(2) & = & \LVentry(3) \\
\LVexit(3) & = & \LVentry(4) \\
\LVexit(4) & = & p_{4,5} \LVentry(5) + p_{4,6} \LVentry(6) \\
\LVexit(5) & = & (1,0) \otimes (1,0) \otimes (1,0) \\
\LVexit(6) & = & (1,0) \otimes (1,0) \otimes (1,0) \\
\end{eqnarray*}
\end{minipage}

And thus the solutions for the probabilistic \LV\ analysis are given by:

\begin{minipage}{8cm}
\begin{eqnarray*}
\LVentry(1) & = & (1,0) \otimes (1,0) \otimes (1,0) \\
\LVentry(2) & = & (0,1) \otimes (1,0) \otimes (1,0) \\
\LVentry(3) & = & 0.25 \cdot (0,1) \otimes (0,1) \otimes (1,0) +  \\ &+&
                                0.75 \cdot (0,1) \otimes (0,1) \otimes (1,0) \\
                     & = & (0,1) \otimes (0,1) \otimes (1,0) \\
\LVentry(4) & = & 0.25 \cdot (0,1) \otimes (1,0) \otimes (1,0) + \\ &+&
                                0.75 \cdot (0,1) \otimes (0,1) \otimes (1,0) \\
\LVentry(5) & = & (0,1) \otimes (1,0) \otimes (1,0) \\
\LVentry(6) & = & (1,0) \otimes (0,1) \otimes (1,0) \\
\end{eqnarray*}
\end{minipage}
\begin{minipage}{6cm}
\begin{eqnarray*}
\LVexit(1) & = &  (0,1) \otimes (1,0) \otimes (1,0) \\
\LVexit(2) & = &  (0,1) \otimes (0,1) \otimes (1,0) \\
\LVexit(3) & = &  0.25 \cdot (0,1) \otimes (1,0) \otimes (1,0) +  \\ &+&
                              0.75 \cdot (0,1) \otimes (0,1) \otimes (1,0) \\
\LVexit(4) & = &  0.25 \cdot (0,1) \otimes (1,0) \otimes (1,0) +  \\ &+&
                              0.75 \cdot (1,0) \otimes (0,1) \otimes (1,0) \\
\LVexit(5) & = & (1,0) \otimes (1,0) \otimes (1,0) \\
\LVexit(6) & = & (1,0) \otimes (1,0) \otimes (1,0) \\
\end{eqnarray*}
\end{minipage}

This means that, for example, at the beginning label $4$, i.e. the test $x>2$ there
are two situations: It can be with probability $\frac{1}{4}$ that only the variable 
$x$ is alive, or with probability $\frac{3}{4}$ both variables $x$ and $y$ are
alive. One could say that $x$ for sure is alive and $y$ only with a 75\% chance.
At the exit of label $4$ the probabilistic \LV\ analysis tells us that with 25\%
chance {\em only} $x$ is alive and with 75\% that $y$ is the {\em only} live
variable. To say that  $x$ is alive with probability $0.25$ and $y$ with $0.75$
probability would be wrong: It is either $x$ or $y$ which is alive and this is reflected
in the joint distributions represented as tensors, which we obtain as solution.
This illustrates that the probabilistic property space cannot be just $\cV(\{x,y,z\})$
but that we need indeed $\cV(\{d,l\})^{\otimes3}$.
\end{example}


\section{Conclusions and Related Work}
\label{Conclusions}

This paper highlights two important aspects of probabilistic program analysis in a 
data-flow style: (i) the use of tensor products in order to represent the correlation 
between a number of variables, and (ii) the use of Probabilistic 
Abstract Interpretation to estimate branch probabilities and to construct probabilistic 
transfer functions. In particular, we argue that static program analysis does not mean 
necessarily considering {\em static branch prediction}. Instead -- by extending
single numbers $p_{\ell,\ell'}$ as branch probabilities to matrices as abstract
branch probabilities $\bP(\ell,\ell)^\#$ -- the PAI framework allows us to express 
dynamic or conditional aspects.

The framework presented here aims in providing a formal basis for speculative 
optimisation. Speculative optimisation \cite{LinEtAl03,BelevantsevEtAl08}\
has been an element of hardware design for some time, in particular to branch prediction 
\cite{McFarling93}\ or for cache optimisation \cite{NicolaescuEtAl06}. More recently, related
ideas have also been discussed in the context of speculative multi-threading
 \cite{BhowmikFranklin04}\ or probabilistic pointer analysis \cite{APLAS07,HungEtAl12}.

The work we have presented in this paper concentrates on the conceptual aspects of probabilistic analysis
and not on optimal realisation of, for example, concrete branch predictors.  Further work 
should however include practical implementations of the presented framework
in order to compare its performance with the large number of  predictors
in existence. Another  research direction concerns the automatic construction 
of abstractions so that  the induced $\bP(\ell,\ell)^\#$ are optimal and maximally predictive.


\bibliographystyle{eptcs}
\bibliography{GandALF13}

\begin{thebibliography}{10}
\providecommand{\bibitemdeclare}[2]{}
\providecommand{\surnamestart}{}
\providecommand{\surnameend}{}
\providecommand{\urlprefix}{Available at }
\providecommand{\url}[1]{\texttt{#1}}
\providecommand{\href}[2]{\texttt{#2}}
\providecommand{\urlalt}[2]{\href{#1}{#2}}
\providecommand{\doi}[1]{doi:\urlalt{http://dx.doi.org/#1}{#1}}
\providecommand{\bibinfo}[2]{#2}

\bibitemdeclare{book}{Dragon2}
\bibitem{Dragon2}
\bibinfo{author}{A.V. \surnamestart Aho\surnameend}, \bibinfo{author}{M.S.
  \surnamestart Lam\surnameend}, \bibinfo{author}{R.~\surnamestart
  Sethi\surnameend} \& \bibinfo{author}{J.D. \surnamestart Ullman\surnameend}
  (\bibinfo{year}{2007}): \emph{\bibinfo{title}{Compilers: {P}rinciples,
  {T}echniques, and {T}ools}}, \bibinfo{edition}{second} edition.
\newblock \bibinfo{publisher}{Pearson Education}.

\bibitemdeclare{article}{BelevantsevEtAl08}
\bibitem{BelevantsevEtAl08}
\bibinfo{author}{A.~A. \surnamestart Belevantsev\surnameend},
  \bibinfo{author}{S.~S. \surnamestart Gaisaryan\surnameend} \&
  \bibinfo{author}{V.~P. \surnamestart Ivannikov\surnameend}
  (\bibinfo{year}{2008}): \emph{\bibinfo{title}{Construction of Speculative
  Optimization Algorithms}}.
\newblock {\sl \bibinfo{journal}{Programming and Computer Software}}
  \bibinfo{volume}{34}(\bibinfo{number}{3}), pp. \bibinfo{pages}{138--153},
  \doi{10.1134/S036176880803002X}.

\bibitemdeclare{book}{BenIsraelGreville03}
\bibitem{BenIsraelGreville03}
\bibinfo{author}{A.~\surnamestart Ben-Israel\surnameend} \&
  \bibinfo{author}{T.N.E. \surnamestart Greville\surnameend}
  (\bibinfo{year}{2003}): \emph{\bibinfo{title}{Generalised Inverses}},
  \bibinfo{edition}{2nd} edition.
\newblock \bibinfo{publisher}{Springer Verlag}.

\bibitemdeclare{article}{BhowmikFranklin04}
\bibitem{BhowmikFranklin04}
\bibinfo{author}{A.~\surnamestart Bhowmik\surnameend} \&
  \bibinfo{author}{M.~\surnamestart Franklin\surnameend}
  (\bibinfo{year}{2004}): \emph{\bibinfo{title}{A General Compiler Framework
  for Speculative Multithreaded Processors}}.
\newblock {\sl \bibinfo{journal}{IEEE Transactions on Parallel and Distributed
  Syststems}} \bibinfo{volume}{15}(\bibinfo{number}{8}), pp.
  \bibinfo{pages}{713--724}, \doi{10.1109/TPDS.2004.26}.

\bibitemdeclare{book}{CampbellMeyer79}
\bibitem{CampbellMeyer79}
\bibinfo{author}{S.L. \surnamestart Campbell\surnameend} \&
  \bibinfo{author}{D.~\surnamestart Meyer\surnameend} (\bibinfo{year}{1979}):
  \emph{\bibinfo{title}{Generalized Inverse of Linear Transformations}}.
\newblock \bibinfo{publisher}{Constable}, \bibinfo{address}{London}.

\bibitemdeclare{inproceedings}{CousotCousot77a}
\bibitem{CousotCousot77a}
\bibinfo{author}{P.~\surnamestart Cousot\surnameend} \&
  \bibinfo{author}{R.~\surnamestart Cousot\surnameend} (\bibinfo{year}{1977}):
  \emph{\bibinfo{title}{Abstract {I}nterpretation: {A} {U}nified {L}attice
  {M}odel for {S}tatic {A}nalysis of {P}rograms by {C}onstruction or
  {A}pproximation of {F}ixpoints}}.
\newblock In: {\sl \bibinfo{booktitle}{POPL'77}}, pp.
  \bibinfo{pages}{238--252}, \doi{10.1145/512950.512973}.

\bibitemdeclare{book}{Deutsch01}
\bibitem{Deutsch01}
\bibinfo{author}{F.~\surnamestart Deutsch\surnameend} (\bibinfo{year}{2001}):
  \emph{\bibinfo{title}{Bet Approximation in Inner Product Spaces}}.
\newblock {\sl \bibinfo{series}{CMS Books in Mathematics}}~\bibinfo{volume}{7},
  \bibinfo{publisher}{Springer Verlag}, \bibinfo{address}{New York --- Berlin},
  \doi{10.1007/978-1-4684-9298-9}.

\bibitemdeclare{incollection}{Festschrift06}
\bibitem{Festschrift06}
\bibinfo{author}{A.~\surnamestart {Di Pierro}\surnameend},
  \bibinfo{author}{C.~\surnamestart Hankin\surnameend} \&
  \bibinfo{author}{H.~\surnamestart Wiklicky\surnameend}
  (\bibinfo{year}{2007}): \emph{\bibinfo{title}{Abstract Interpretation for
  Worst and Average Case Analysis}}.
\newblock In: {\sl \bibinfo{booktitle}{Program Analysis and Compilation, Theory
  and Practice}}, {\sl \bibinfo{series}{LNCS}} \bibinfo{volume}{4444},
  \bibinfo{publisher}{Springer Verlag}, pp. \bibinfo{pages}{160--174},
  \doi{10.1007/978-3-540-71322-7\_8}.

\bibitemdeclare{inproceedings}{APLAS07}
\bibitem{APLAS07}
\bibinfo{author}{A.~\surnamestart {Di Pierro}\surnameend},
  \bibinfo{author}{C.~\surnamestart Hankin\surnameend} \&
  \bibinfo{author}{H.~\surnamestart Wiklicky\surnameend}
  (\bibinfo{year}{2007}): \emph{\bibinfo{title}{A Systematic Approach to
  Probabilistic Pointer Analysis}}.
\newblock In \bibinfo{editor}{Z.~\surnamestart Shao\surnameend}, editor: {\sl
  \bibinfo{booktitle}{Proceedings of APLAS'07}}, {\sl \bibinfo{series}{LNCS}}
  \bibinfo{volume}{4807}, \bibinfo{publisher}{Springer Verlag}, pp.
  \bibinfo{pages}{335--350}, \doi{10.1007/978-3-540-76637-7\_23}.

\bibitemdeclare{incollection}{Bertinoro10}
\bibitem{Bertinoro10}
\bibinfo{author}{A.~\surnamestart {Di Pierro}\surnameend},
  \bibinfo{author}{C.~\surnamestart Hankin\surnameend} \&
  \bibinfo{author}{H.~\surnamestart Wiklicky\surnameend}
  (\bibinfo{year}{2010}): \emph{\bibinfo{title}{Probabilistic Semantics and
  Analysis}}.
\newblock In: {\sl \bibinfo{booktitle}{Formal Methods for Quantitative Aspects
  of Programming Languages}}, {\sl \bibinfo{series}{LNCS}}
  \bibinfo{volume}{6155}, \bibinfo{publisher}{Springer Verlag}, pp.
  \bibinfo{pages}{1--42}, \doi{10.1007/978-3-642-13678-8\_1}.

\bibitemdeclare{inproceedings}{QAPL08}
\bibitem{QAPL08}
\bibinfo{author}{A.~\surnamestart {Di Pierro}\surnameend},
  \bibinfo{author}{P.~\surnamestart Sotin\surnameend} \&
  \bibinfo{author}{H.~\surnamestart Wiklicky\surnameend}
  (\bibinfo{year}{2008}): \emph{\bibinfo{title}{Relational Analysis and
  Precision via Probabilistic Abstract Interpretation}}.
\newblock In \bibinfo{editor}{C.~\surnamestart Baier\surnameend} \&
  \bibinfo{editor}{A.~\surnamestart Aldini\surnameend}, editors: {\sl
  \bibinfo{booktitle}{Proceedings of QAPL'08}}, \bibinfo{series}{Electronic
  Notes in Theoretical Computer Science}, \bibinfo{publisher}{Elsevier}, pp.
  \bibinfo{pages}{23--42}, \doi{10.1016/j.entcs.2008.11.017}.

\bibitemdeclare{inproceedings}{PPDP00}
\bibitem{PPDP00}
\bibinfo{author}{A.~\surnamestart {Di Pierro}\surnameend} \&
  \bibinfo{author}{H.~\surnamestart Wiklicky\surnameend}
  (\bibinfo{year}{2000}): \emph{\bibinfo{title}{{C}oncurrent {C}onstraint
  {P}rogramming: Towards {P}robabilistic {A}bstract {I}nterpretation}}.
\newblock In: {\sl \bibinfo{booktitle}{PPDP'00}}, pp.
  \bibinfo{pages}{127--138}, \doi{10.1145/351268.351284}.

\bibitemdeclare{article}{HungEtAl12}
\bibitem{HungEtAl12}
\bibinfo{author}{M.-Y. \surnamestart Hung\surnameend}, \bibinfo{author}{P.-S.
  \surnamestart Chen\surnameend}, \bibinfo{author}{Y-S. \surnamestart
  Hwang\surnameend}, \bibinfo{author}{R.~D.-C. \surnamestart Ju\surnameend} \&
  \bibinfo{author}{J.~K. \surnamestart Lee\surnameend} (\bibinfo{year}{2012}):
  \emph{\bibinfo{title}{Support of Probabilistic Pointer Analysis in the SSA
  Form}}.
\newblock {\sl \bibinfo{journal}{IEEE Transactions on Parallel Distributed
  Syststems}} \bibinfo{volume}{23}(\bibinfo{number}{12}), pp.
  \bibinfo{pages}{2366--2379}, \doi{10.1109/TPDS.2012.73}.

\bibitemdeclare{inproceedings}{PRISM04}
\bibitem{PRISM04}
\bibinfo{author}{M.Z. \surnamestart Kwiatkowska\surnameend},
  \bibinfo{author}{G.~\surnamestart Norman\surnameend} \&
  \bibinfo{author}{D.~\surnamestart Parker\surnameend} (\bibinfo{year}{2004}):
  \emph{\bibinfo{title}{PRISM 2.0: A Tool for Probabilistic Model Checking}}.
\newblock In: {\sl \bibinfo{booktitle}{International Conference on Quantitative
  Evaluation of Systems (QEST 2004)}}, \bibinfo{publisher}{IEEE Computer
  Society}, pp. \bibinfo{pages}{322--323}, \doi{10.1109/QEST.2004.10016}.

\bibitemdeclare{inproceedings}{LinEtAl03}
\bibitem{LinEtAl03}
\bibinfo{author}{J.~\surnamestart Lin\surnameend},
  \bibinfo{author}{T.~\surnamestart Chen\surnameend}, \bibinfo{author}{W.-C.
  \surnamestart Hsu\surnameend}, \bibinfo{author}{P.-C. \surnamestart
  Yew\surnameend}, \bibinfo{author}{R.~D.-C. \surnamestart Ju\surnameend},
  \bibinfo{author}{T.-F. \surnamestart Ngai\surnameend} \&
  \bibinfo{author}{S.~\surnamestart Chan\surnameend} (\bibinfo{year}{2003}):
  \emph{\bibinfo{title}{A compiler framework for speculative analysis and
  optimizations}}.
\newblock In: {\sl \bibinfo{booktitle}{Proceedings Conference on Programming
  Language Design and Implementation (PLDI)}}, pp. \bibinfo{pages}{289--299},
  \doi{10.1145/781131.781164}.

\bibitemdeclare{techreport}{McFarling93}
\bibitem{McFarling93}
\bibinfo{author}{S.~\surnamestart McFarling\surnameend} (\bibinfo{year}{1993}):
  \emph{\bibinfo{title}{Combining Branch Predictors}}.
\newblock \bibinfo{type}{Technical Report} \bibinfo{number}{WLR TN-36},
  \bibinfo{institution}{Digital}.

\bibitemdeclare{inproceedings}{NicolaescuEtAl06}
\bibitem{NicolaescuEtAl06}
\bibinfo{author}{D.~\surnamestart Nicolaescu\surnameend},
  \bibinfo{author}{B.~\surnamestart Salamat\surnameend} \&
  \bibinfo{author}{A.V. \surnamestart Veidenbaum\surnameend}
  (\bibinfo{year}{2006}): \emph{\bibinfo{title}{Fast Speculative Address
  Generation and Way Caching for Reducing L1 Data Cache Energy}}.
\newblock In: {\sl \bibinfo{booktitle}{Proceedings of the 24th International
  Conference on Computer Design (ICCD 2006)}}, \bibinfo{publisher}{IEEE}, pp.
  \bibinfo{pages}{101--107}, \doi{10.1109/ICCD.2006.4380801}.

\bibitemdeclare{book}{NielsonEtAl99}
\bibitem{NielsonEtAl99}
\bibinfo{author}{F.~\surnamestart Nielson\surnameend}, \bibinfo{author}{H.~Riis
  \surnamestart Nielson\surnameend} \& \bibinfo{author}{C.~\surnamestart
  Hankin\surnameend} (\bibinfo{year}{1999}): \emph{\bibinfo{title}{{P}rinciples
  of {P}rogram {A}nalysis}}.
\newblock \bibinfo{publisher}{Springer Verlag}, \bibinfo{address}{Berlin --
  Heidelberg}, \doi{10.1007/978-3-662-03811-6}.

\bibitemdeclare{book}{Roman05}
\bibitem{Roman05}
\bibinfo{author}{S.~\surnamestart Roman\surnameend} (\bibinfo{year}{2005}):
  \emph{\bibinfo{title}{Advanced Linear Algebra}}, \bibinfo{edition}{2nd}
  edition.
\newblock \bibinfo{publisher}{Springer Verlag}.

\bibitemdeclare{article}{StylesLuk04}
\bibitem{StylesLuk04}
\bibinfo{author}{H.~\surnamestart Styles\surnameend} \&
  \bibinfo{author}{W.~\surnamestart Luk\surnameend} (\bibinfo{year}{2004}):
  \emph{\bibinfo{title}{Exploiting Program Branch Probabilities in Hardware
  Compilation}}.
\newblock {\sl \bibinfo{journal}{IEEE Transaction on Computers}}
  \bibinfo{volume}{53}(\bibinfo{number}{11}), pp. \bibinfo{pages}{1408--1419},
  \doi{10.1109/TC.2004.96}.

\end{thebibliography}

\appendix
\section*{Appendix}

For completeness, we present here the abstract transfer functions
in the probabilistic analysis of Example~\ref{prunning}.

\footnotesize
\[
\bF_1^\# =
\left(
\begin{array}{cccccccccccccccc}
\frac{1}{2} & 0 & 0 & 0 & \frac{1}{2} & 0 & 0 & 0 & 0 & 0 & 0 & 0 & 0 & 0 & 0 & 0 \\ 
0 & \frac{1}{2} & 0 & 0 & 0 & \frac{1}{2} & 0 & 0 & 0 & 0 & 0 & 0 & 0 & 0 & 0 & 0 \\ 
0 & 0 & \frac{1}{2} & 0 & 0 & 0 & \frac{1}{2} & 0 & 0 & 0 & 0 & 0 & 0 & 0 & 0 & 0 \\ 
0 & 0 & 0 & \frac{1}{2} & 0 & 0 & 0 & \frac{1}{2} & 0 & 0 & 0 & 0 & 0 & 0 & 0 & 0 \\ 
\frac{1}{2} & 0 & 0 & 0 & \frac{1}{2} & 0 & 0 & 0 & 0 & 0 & 0 & 0 & 0 & 0 & 0 & 0 \\ 
0 & \frac{1}{2} & 0 & 0 & 0 & \frac{1}{2} & 0 & 0 & 0 & 0 & 0 & 0 & 0 & 0 & 0 & 0 \\ 
0 & 0 & \frac{1}{2} & 0 & 0 & 0 & \frac{1}{2} & 0 & 0 & 0 & 0 & 0 & 0 & 0 & 0 & 0 \\ 
0 & 0 & 0 & \frac{1}{2} & 0 & 0 & 0 & \frac{1}{2} & 0 & 0 & 0 & 0 & 0 & 0 & 0 & 0 \\ 
\frac{1}{2} & 0 & 0 & 0 & \frac{1}{2} & 0 & 0 & 0 & 0 & 0 & 0 & 0 & 0 & 0 & 0 & 0 \\ 
0 & \frac{1}{2} & 0 & 0 & 0 & \frac{1}{2} & 0 & 0 & 0 & 0 & 0 & 0 & 0 & 0 & 0 & 0 \\ 
0 & 0 & \frac{1}{2} & 0 & 0 & 0 & \frac{1}{2} & 0 & 0 & 0 & 0 & 0 & 0 & 0 & 0 & 0 \\ 
0 & 0 & 0 & \frac{1}{2} & 0 & 0 & 0 & \frac{1}{2} & 0 & 0 & 0 & 0 & 0 & 0 & 0 & 0 \\ 
\frac{1}{2} & 0 & 0 & 0 & \frac{1}{2} & 0 & 0 & 0 & 0 & 0 & 0 & 0 & 0 & 0 & 0 & 0 \\ 
0 & \frac{1}{2} & 0 & 0 & 0 & \frac{1}{2} & 0 & 0 & 0 & 0 & 0 & 0 & 0 & 0 & 0 & 0 \\ 
0 & 0 & \frac{1}{2} & 0 & 0 & 0 & \frac{1}{2} & 0 & 0 & 0 & 0 & 0 & 0 & 0 & 0 & 0 \\ 
0 & 0 & 0 & \frac{1}{2} & 0 & 0 & 0 & \frac{1}{2} & 0 & 0 & 0 & 0 & 0 & 0 & 0 & 0 \\ 
\end{array}
\right)
\]
\[
\bF_2^\# =
\left(
\begin{array}{cccccccccccccccc}
\frac{1}{4} & \frac{1}{4} & \frac{1}{4} & \frac{1}{4} & 0 & 0 & 0 & 0 & 0 & 0 & 0 & 0 & 0 & 0 & 0 & 0 \\ 
\frac{1}{4} & \frac{1}{4} & \frac{1}{4} & \frac{1}{4} & 0 & 0 & 0 & 0 & 0 & 0 & 0 & 0 & 0 & 0 & 0 & 0 \\ 
\frac{1}{4} & \frac{1}{4} & \frac{1}{4} & \frac{1}{4} & 0 & 0 & 0 & 0 & 0 & 0 & 0 & 0 & 0 & 0 & 0 & 0 \\ 
\frac{1}{4} & \frac{1}{4} & \frac{1}{4} & \frac{1}{4} & 0 & 0 & 0 & 0 & 0 & 0 & 0 & 0 & 0 & 0 & 0 & 0 \\ 
0 & 0 & 0 & 0 & \frac{1}{4} & \frac{1}{4} & \frac{1}{4} & \frac{1}{4} & 0 & 0 & 0 & 0 & 0 & 0 & 0 & 0 \\ 
0 & 0 & 0 & 0 & \frac{1}{4} & \frac{1}{4} & \frac{1}{4} & \frac{1}{4} & 0 & 0 & 0 & 0 & 0 & 0 & 0 & 0 \\ 
0 & 0 & 0 & 0 & \frac{1}{4} & \frac{1}{4} & \frac{1}{4} & \frac{1}{4} & 0 & 0 & 0 & 0 & 0 & 0 & 0 & 0 \\ 
0 & 0 & 0 & 0 & \frac{1}{4} & \frac{1}{4} & \frac{1}{4} & \frac{1}{4} & 0 & 0 & 0 & 0 & 0 & 0 & 0 & 0 \\ 
0 & 0 & 0 & 0 & 0 & 0 & 0 & 0 & \frac{1}{4} & \frac{1}{4} & \frac{1}{4} & \frac{1}{4} & 0 & 0 & 0 & 0 \\ 
0 & 0 & 0 & 0 & 0 & 0 & 0 & 0 & \frac{1}{4} & \frac{1}{4} & \frac{1}{4} & \frac{1}{4} & 0 & 0 & 0 & 0 \\ 
0 & 0 & 0 & 0 & 0 & 0 & 0 & 0 & \frac{1}{4} & \frac{1}{4} & \frac{1}{4} & \frac{1}{4} & 0 & 0 & 0 & 0 \\ 
0 & 0 & 0 & 0 & 0 & 0 & 0 & 0 & \frac{1}{4} & \frac{1}{4} & \frac{1}{4} & \frac{1}{4} & 0 & 0 & 0 & 0 \\ 
0 & 0 & 0 & 0 & 0 & 0 & 0 & 0 & 0 & 0 & 0 & 0 & \frac{1}{4} & \frac{1}{4} & \frac{1}{4} & \frac{1}{4} \\ 
0 & 0 & 0 & 0 & 0 & 0 & 0 & 0 & 0 & 0 & 0 & 0 & \frac{1}{4} & \frac{1}{4} & \frac{1}{4} & \frac{1}{4} \\ 
0 & 0 & 0 & 0 & 0 & 0 & 0 & 0 & 0 & 0 & 0 & 0 & \frac{1}{4} & \frac{1}{4} & \frac{1}{4} & \frac{1}{4} \\ 
0 & 0 & 0 & 0 & 0 & 0 & 0 & 0 & 0 & 0 & 0 & 0 & \frac{1}{4} & \frac{1}{4} & \frac{1}{4} & \frac{1}{4} \\ 
\end{array}
\right)
\]
\[
\bF_3^\# =
\left(
\begin{array}{cccccccccccccccc}
1 & 0 & 0 & 0 & 0 & 0 & 0 & 0 & 0 & 0 & 0 & 0 & 0 & 0 & 0 & 0 \\ 
0 & 0 & 0 & 0 & 0 & 1 & 0 & 0 & 0 & 0 & 0 & 0 & 0 & 0 & 0 & 0 \\ 
0 & 0 & 0 & 0 & 0 & 0 & 0 & 0 & 0 & 0 & 1 & 0 & 0 & 0 & 0 & 0 \\ 
0 & 0 & 0 & 0 & 0 & 0 & 0 & 0 & 0 & 0 & 0 & 0 & 0 & 0 & 0 & 1 \\ 
0 & 0 & 0 & 0 & 1 & 0 & 0 & 0 & 0 & 0 & 0 & 0 & 0 & 0 & 0 & 0 \\ 
0 & 0 & 0 & 0 & 0 & 0 & 0 & 0 & 0 & 1 & 0 & 0 & 0 & 0 & 0 & 0 \\ 
0 & 0 & 0 & 0 & 0 & 0 & 0 & 0 & 0 & 0 & 0 & 0 & 0 & 0 & 1 & 0 \\ 
0 & 0 & 0 & 1 & 0 & 0 & 0 & 0 & 0 & 0 & 0 & 0 & 0 & 0 & 0 & 0 \\ 
0 & 0 & 0 & 0 & 0 & 0 & 0 & 0 & 1 & 0 & 0 & 0 & 0 & 0 & 0 & 0 \\ 
0 & 0 & 0 & 0 & 0 & 0 & 0 & 0 & 0 & 0 & 0 & 0 & 0 & 1 & 0 & 0 \\ 
0 & 0 & 1 & 0 & 0 & 0 & 0 & 0 & 0 & 0 & 0 & 0 & 0 & 0 & 0 & 0 \\ 
0 & 0 & 0 & 0 & 0 & 0 & 0 & 1 & 0 & 0 & 0 & 0 & 0 & 0 & 0 & 0 \\ 
0 & 0 & 0 & 0 & 0 & 0 & 0 & 0 & 0 & 0 & 0 & 0 & 1 & 0 & 0 & 0 \\ 
0 & 1 & 0 & 0 & 0 & 0 & 0 & 0 & 0 & 0 & 0 & 0 & 0 & 0 & 0 & 0 \\ 
0 & 0 & 0 & 0 & 0 & 0 & 1 & 0 & 0 & 0 & 0 & 0 & 0 & 0 & 0 & 0 \\ 
0 & 0 & 0 & 0 & 0 & 0 & 0 & 0 & 0 & 0 & 0 & 1 & 0 & 0 & 0 & 0 \\ 
\end{array}
\right)
\]
\[
\bP_4^\# =
\left(
\begin{array}{cccccccccccccccc}
1 & 0 & 0 & 0 & 0 & 0 & 0 & 0 & 0 & 0 & 0 & 0 & 0 & 0 & 0 & 0 \\ 
0 & 1 & 0 & 0 & 0 & 0 & 0 & 0 & 0 & 0 & 0 & 0 & 0 & 0 & 0 & 0 \\ 
0 & 0 & 1 & 0 & 0 & 0 & 0 & 0 & 0 & 0 & 0 & 0 & 0 & 0 & 0 & 0 \\ 
0 & 0 & 0 & 1 & 0 & 0 & 0 & 0 & 0 & 0 & 0 & 0 & 0 & 0 & 0 & 0 \\ 
0 & 0 & 0 & 0 & 1 & 0 & 0 & 0 & 0 & 0 & 0 & 0 & 0 & 0 & 0 & 0 \\ 
0 & 0 & 0 & 0 & 0 & 1 & 0 & 0 & 0 & 0 & 0 & 0 & 0 & 0 & 0 & 0 \\ 
0 & 0 & 0 & 0 & 0 & 0 & 1 & 0 & 0 & 0 & 0 & 0 & 0 & 0 & 0 & 0 \\ 
0 & 0 & 0 & 0 & 0 & 0 & 0 & 1 & 0 & 0 & 0 & 0 & 0 & 0 & 0 & 0 \\ 
0 & 0 & 0 & 0 & 0 & 0 & 0 & 0 & 1 & 0 & 0 & 0 & 0 & 0 & 0 & 0 \\ 
0 & 0 & 0 & 0 & 0 & 0 & 0 & 0 & 0 & 1 & 0 & 0 & 0 & 0 & 0 & 0 \\ 
0 & 0 & 0 & 0 & 0 & 0 & 0 & 0 & 0 & 0 & 1 & 0 & 0 & 0 & 0 & 0 \\ 
0 & 0 & 0 & 0 & 0 & 0 & 0 & 0 & 0 & 0 & 0 & 1 & 0 & 0 & 0 & 0 \\ 
0 & 0 & 0 & 0 & 0 & 0 & 0 & 0 & 0 & 0 & 0 & 0 & 0 & 0 & 0 & 0 \\ 
0 & 0 & 0 & 0 & 0 & 0 & 0 & 0 & 0 & 0 & 0 & 0 & 0 & 0 & 0 & 0 \\ 
0 & 0 & 0 & 0 & 0 & 0 & 0 & 0 & 0 & 0 & 0 & 0 & 0 & 0 & 0 & 0 \\ 
0 & 0 & 0 & 0 & 0 & 0 & 0 & 0 & 0 & 0 & 0 & 0 & 0 & 0 & 0 & 0 \\
\end{array}
\right)
\]

\end{document}